\documentclass[a4paper,11pt]{article}
\usepackage{graphicx,amssymb,bm,latexsym,color,epsf,cite,ulem,verbatim}

\usepackage{hyperref}

\makeatletter
\@addtoreset{equation}{section}

\makeatother
\newcommand{\be}{\begin{equation}}
\newcommand{\ee}{\end{equation}}
\newcommand{\bea}{\begin{eqnarray}}
\newcommand{\eea}{\end{eqnarray}}
\newcommand{\vs}[1]{\vspace{#1 mm}}

\renewcommand{\a}{\alpha}
\renewcommand{\b}{\beta}

\renewcommand{\d}{\delta}

\newcommand{\vp}{\varphi}

\newcommand{\pa}{\partial}

\newcommand{\nn}{\nonumber\\}

\newcommand{\cD}{{\cal D}}

\newcommand{\cL}{{\cal L}}

\newcommand{\tr}{{\rm tr}}
\newcommand{\Tr}{{\rm Tr}}

\newcommand{\diff}{\mathcal{D}if\!f(M_R,M_L)}
\newcommand{\diffl}{\mathcal{D}if\!f M_L}
\newcommand{\diffr}{\mathcal{D}if\!f M_R}
\newcommand{\diffM}{\mathcal{D}if\!f(M)}

\begin{document}

\begin{center}
{\Large\bf Dynamical diffeomorphisms}

\vs{8}

{\large
Renata Ferrero\footnote{e-mail address: rferrero@uni-mainz.de}
and Roberto Percacci\footnote{e-mail address: percacci@sissa.it}$^{,3}$
} \\
\vs{8}
$^1${\textit{Institut f\"{u}r Physik (THEP), Johannes Gutenberg - Universit\"{a}t Mainz,
\\Staudingerweg 7, {55128 Mainz, Germany}}

$^2${\textit{International School for Advanced Studies, via Bonomea 265, 34136 Trieste, Italy}}

$^3${\textit{INFN, Sezione di Trieste, Italy}}
}
\end{center}
\vs{1}

\setcounter{footnote}{0} 

\begin{abstract}
We construct a general effective dynamics for diffeomorphisms of spacetime, in a fixed external metric.
Though related to familiar models of scalar fields
as coordinates, our models have subtly different properties,
both at kinematical and dynamical level.
The energy-momentum tensor consists of two
independently conserved parts.
The background solution is the identity diffeomorphism
and the energy-momentum tensor of this solution gives 
rise to an effective cosmological constant.
\end{abstract}

\normalsize
\section{Introduction}

In General Relativity, spacetime diffeomorphisms play the role
of active gauge transformations,
while coordinate transformations are usually viewed as
passive gauge transformations.
In this paper we shall discuss instead a possible role
of diffeomorphisms as dynamical variables,
a point of view that is closer to their applications
in hydrodynamics and elasticity.
We have in mind two main applications:
the problem of dark energy in cosmology
and the problem of observables in General Relativity.
In this introduction we shall review these motivations.

Modelling dark energy is important
both for early and late cosmology.
A term in the action proportional to the volume of spacetime 
is the simplest explanation, but it has drawbacks.
For this reason a more dynamical origin is often preferred.
See \cite{Amendola:2015ksp} for a review of
many possible alternatives.
The most popular models are based on the potential
of dynamical scalars.
There are also many models of scalar fields
with derivative interactions.
The models we will discuss in this paper
can be seen as a special subclass of the latter, 
where the fields are restricted kinematically to
be diffeomorphisms of spacetime.
The idea of using the dynamics of coordinates or diffeomorphisms
to generate an effective cosmological constant goes back at least
to \cite{omero,GellMann:1984sj,GellMann:1985if}, where it was used to induce spontaneous compactification
of certain directions in higher-dimensional theories.
More recently, it has been used extensively 
in the literature on massive gravity \cite{Hinterbichler:2011tt}.
A mass term for the graviton breaks diffeomorphism symmetry,
but the theory can still be written in a diffeomorphism-invariant way
by introducing four ``St\"{u}ckelberg'' fields,
much in the same way as massive QED can be written
in a gauge-invariant way by introducing one real scalar field
\cite{Stueckelberg:1957zz}.
In an influential paper, Arkani-Hamed, Georgi and Schwartz
\cite{ArkaniHamed:2002sp} 
have constructed the effective field theory of the four 
Goldstone bosons that are used to restore diffeomorphism invariance 
in massive gravity.
The theory is formulated in terms of two separate ``sites'',
that can be viewed as two copies of spacetime,
each endowed with a separate diffeomorphism invariance, 
and a field linking the two sites.
As with all Goldstone bosons,
the Lagrangian of the fluctuation of the linking field 
is shift-invariant and therefore contains only derivative couplings.

The major issue with this idea is that one of the four scalars
(namely the one associated to the time coordinate) is a ghost.
\footnote{In the models of \cite{omero,GellMann:1984sj,GellMann:1985if}, the problem of ghosts did not arise
because the compactified dimensions are spacelike.}
This is because the target space of the scalar fields has
a Minkowskian metric.
In the original model, the ghost 
starts propagating at energy scales higher than
$(1000\textrm{km})^{-1}$
\cite{Boulware:1973my,Creminelli:2005qk},
leading to strong conflict with observations.
This problem can be circumvented in two different ways.
The first goes back to the fully non-linear version of the
massive Fierz-Pauli theory, constructed
as a bi-metric theory with a specific potential
\cite{deRham:2010kj,deRham:2011rn,Hassan:2011hr,Hassan:2011vm,Hassan:2011ea,deRham:2016plk}.
This theory can be made diffeomorphism-invariant
by adding suitable scalar fields and is ghost-free.
{\it In vacuo} it is reliable up to energies of order $\sim \textrm{mm}^{-1}$; above this scale, it becomes invalid due to strong coupling effects. In the presence of sources the situation is much worse.

The second is to construct a ghost-free theory 
reducing diffeomorphism invariance to
foliation-preserving diffeomorphisms
\cite{Rubakov:2004eb, Dubovsky:2004sg, Comelli:2013txa, Comelli:2013paa}.
This is perhaps not too high a price to be paid,
since in the cosmological context a preferred foliation is
singled out anyway.
\footnote{We recall that the same symmetry reduction is
also present in Ho\v{r}ava-Lifshitz gravity
\cite{Horava:2009uw},
and is sufficient to reconcile ghost freedom with
perturbative renormalizability.}
The resulting models can be interpreted as describing the dynamics of a medium filling the Universe. 
See \cite{Endlich:2012pz,Comelli:2013tja, Comelli:2014xga, Ballesteros:2016gwc, Celoria:2017bbh, Celoria:2017idi, Celoria:2019oiu} for later developments and applications to accelerating cosmological models.

Restricting ourselves to the covariant models, 
the main building block of the action is the matrix
\begin{eqnarray}
B^\mu{}_\nu=
g^{\mu\rho}(\varphi^*h)_{\rho\nu}
=g^{\mu\rho}\pa_\rho\varphi^\alpha 
\pa_\nu\varphi^\beta
h_{\alpha\beta}(\varphi)
\ ,
\label{B1}
\end{eqnarray}
where $\varphi^\alpha$ are the scalar fields,
$g_{\mu\nu}$ is the spacetime metric,
$h_{\alpha\beta}(\varphi)$ is the target space metric
(usually assumed flat)
and $\varphi^*h$ is the pullback of $h$ by $\varphi$.
A necessary condition for $\varphi$ to be a diffeomorphisms
is that $\partial_\mu\varphi^\alpha$ be nondegenerate everywhere.
The transformations of $B^\mu{}_\nu$ are the same as one would have in an ordinary nonlinear sigma model in curved space,
and diffeomorphism-invariant actions can be constructed
by taking functions of traces of this matrix.
\footnote{Obviously, diffeomorphism invariance requires that the metric be transformed.}
For example, the simplest action is the action for harmonic maps
\be
S=-\frac12 f^2\int d^4x\sqrt{g}\,\tr B\ ,
\label{harmonic}
\ee
where $f$ is a constant with dimension of mass.
In the non-covariant models with only foliation-preserving
diffeomorphism invariance,
the indices $\alpha$, $\beta$
run only from $1$ to $d-1$ and are associated
with the space coordinates.

In these models it is generally assumed that the topology of
spacetime is $\mathbb{R}^4$.
This is in line with the standard mathematical
definition of coordinates as maps from a manifold to $\mathbb{R}^4$.
How could the models be extended to topologically nontrivial spacetimes,
for example cosmology with compact spatial sections?
This seems hard as long as one sticks to the notion
of coordinates as dynamical fields,
because coordinates are local constructions
and trying to extend a coordinate system on, e.g., a sphere,
will necessarily lead to singularities.
It is then better to jump from the passive view of
gauge transformations to the active one, namely diffeomorphisms.
Unlike coordinate transformations, diffeomorphisms have
a global definition and one may expect that it is possible
to discuss the dynamics of diffeomorphisms globally.
This is what we do in the present paper.

The other general motivation for our study is
the absence of local observables in (quantum) gravity 
\cite{Rovelli:1990ph, Rovelli:2004tv, Torre:1993fq}.
This is in apparent blatant contradiction to the fact that we
routinely perform measurements of local fields:
for example, one can measure the components of the Riemann tensor
near the surface of the earth.
Upon reflection, one immediately understands that such measurements
are made possible by the existence of matter.
For example, the Ricci scalar $R$ at a point $x$ is not diffeomorphism
invariant: under a diffeomorphism $\psi$, the scalar is transformed
by pullback $R\mapsto \psi^*R$,
so $R(x)\mapsto\psi^*R(x)=R(\psi(x))$.
So, the Ricci scalar is not an observable.
However, if $X$ denotes the (spacetime) position of a particle,
a diffeomorphism will map $X\mapsto \psi^{-1}(X)$.
Thus $R(X)$, the Ricci scalar at the position of the particle,
is diffeomorphism invariant, and hence observable.
With a bunch of infinitesimally close particles one can 
construct a local frame at $X$ and give physical meaning
to the components of any tensor at $X$.
A sufficiently dense dust of such particles,
which in the limit becomes a continuous fluid,
can be used to set up a physical coordinate system
on some patch of spacetime, and hence give physical meaning
to continuous tensor fields.
This is standard procedure in cosmology (see for example \cite{Brown:1994py, Kuchar:1990vy, Giesel:2007wi} or \cite{Tambornino:2011vg} for a review).

All measurements of tensor or scalar fields,
referring to such ``physical'' coordinate systems,
can be thought of as measurements of relational observables.
This is abundantly discussed in the literature \cite{DeWitt:1962cg,Rovelli:1990ph, Rovelli:1990pi, Rovelli:2004tv, Dittrich:2004cb, Dittrich:2005kc,Westman:2007yx,Gielen:2018fqv,Hoehn:2018whn,Hoehn:2019owq}. 
The matter content in the Universe
represents a reference system or reference medium with respect 
to which the points of the target space are defined. 
The question arises: what is the dynamics of these fields?
In many applications the ``coordinate infrastructure''
is sufficiently rarefied that its effect on the geometry can
be safely ignored.
In general, however, this is not the case and the backreaction of
the ``coordinate infrastructure'' on the geometry 
must be accounted for, as for example in cosmology.
In particular, in the classical theory one can assume that
the coordinate fluid is as ``thin'' as one wants,
but not in the quantum theory.
Giddings et al. discuss relational approaches to locality based on diffeomorphism-invariant nonlocal operators \cite{Giddings:2015lla, Donnelly:2016rvo}. In \cite{Giddings:2005id} they have discussed absolute limits.
A simplified version of their argument would go as follows:
in order to be able to distinguish points in spacetime with a 
resolution $\ell$, one needs to fill space with radiation
of wavelength $\lambda<\ell$, and there must be at least one 
quantum per volume $\ell^3$.
This absolutely minimal coordinate infrastructure would
have an energy density $\ell^{-4}$.

One can draw from these remarks two general conclusions.
The first is that in order to meaningfully talk about
local observables in the theory of gravity
one has to include a form of matter
as a coordinate system, and that this will always
back-react onto the geometry. 
Pure gravity is an unphysical abstraction.
The second is that at least some of the conceptual
issues that arise in (quantum) gravity, are in fact the
result of this unphysical abstraction, namely neglecting the
existence of matter.
This is another motivation for models of gravity 
endowed with physical coordinates.
In the hamiltonian framework, this is called ``deparametrization''.
The most popular models of this type contain some form of dust
\cite{Brown:1994py} or phantom energy \cite{Thiemann:2006up}.
There is a close analogue of this in gauge theories,
where gauge-invariant observables can be constructed
by suitably dressing local operators \cite{Frohlich:1981yi,Maas:2017xzh}.
In the case of gravity this has been discussed to some extent in \cite{Maas:2019eux}.
Whether our models can be useful in the construction of relational
observables of this type is a question 
that we shall briefly return to in the conclusion.

After these motivations, 
we remark that treating diffeomorphisms as dynamical variables
gives rise to a theory with rather different features from any other.
Indeed, ordinary matter propagating in spacetime can be represented
either by maps into spacetime, such as the worldlines
of point particles or the worldsheets of strings,
or maps from spacetime into some other ``internal'' space,
as is the case with all the usual matter fields.
Here we are in a peculiar situation where
the ``matter'' field is both a map {\it on} spacetime
and {\it in} spacetime.
This gives rise to significant differences
compared to the dynamics of ordinary matter fields,
and also to the dynamics of coordinates,
regarded as scalar fields on spacetime.
At the level of kinematics, we replace the target space
$\mathbb{R}^4$ by spacetime itself.
\footnote{
There is an analog of this in the St\"uckelberg treatment
of massive QED.
If one takes topology properly into account,
the St\"uckelberg scalar is not an 
ordinary real-valued scalar, but has to be identified modulo $2\pi$.}
As we shall see, quite aside from the topological issues,
this subtly changes the invariance group.
In writing the action, the main novelty of our treatment
is that in (\ref{B1}) we put $h(\varphi)=g(\varphi)$.
This is a rather significant difference, because there is
a new dependence on the metric that affects the
definition of the energy-momentum tensor of the scalars,
and has further consequences on the
relation between the equations of motion and
energy-momentum conservation.
In particular, the energy-momentum tensor consists of two pieces
that are separately conserved.

\bigskip
This paper is organized as follows.
In Section 2 we write a class of actions
for diffeomorphisms of a manifold to itself.
As discussed above,
these differ from the models that have been considerd previously because the metrics in the domain and in the target space
are one and the same (i.e., $g=h$).
We derive the equations of motion (EOMs)
and show that the energy-momentum (EM) tensor contains
new terms that had not been considered previously, 
to the best of our knowledge.
Interestingly, these new terms are conserved
independently of the rest.
We show that the equation of motion of the scalar
is generically equivalent to the conservation of 
the EM tensor, but not always.
In Section 3 we discuss the identity solution as a model
for dark energy and its stability.
In Section 4 we return to the motivations given above
and discuss the extent to which the models may
provide satisfactory answers.
Appendix A contains a detailed proof of the diffeomorphism
invariance of the action.
In Appendix B we show that models where the domain and target 
space are not identified (and in particular $h\not= g$)
have different properties from those discussed in the main text.
However, the latter can be obtained from the former
under some additional conditions.

\section{The models}

The models we shall discuss are very similar to nonlinear
sigma models, except that the domain and the target space
are the same manifold $M$.
In spite of this identification, we still need to distinguish 
two types of geometric objects:
tensors evaluated at a point $x$ and tensors evaluated
at $\varphi(x)$.
In order to better keep track of the difference,
in component formulas, we shall use letters from the
middle greek alphabet for the former and letters
from the beginning of the greek alphabet for the latter.
The pullback transforms tensors at $\varphi(x)$ to tensors at $x$,
for example given a covariant vector $\omega$ (a one-form),
its pullback is
\be
(\varphi^*\omega)_\mu=
\partial_\mu\varphi^\alpha\omega_\alpha(\varphi)
\ee
and
\be
(\varphi^* g)_{\mu\nu}=
\partial_\mu\varphi^\alpha\partial_\nu\varphi^\beta g_{\alpha\beta}(\varphi)
\ee
is the pullback of the metric.
We will always use this notation:
a tensor like $\omega$ without argument is understood to be evaluated
at some point $x$, whereas $\omega(\varphi)$ has to be
understood as a tensor evaluated at $\varphi(x)$.

In this setting the use of covariant derivatives requires a little 
explanation.
The covariant derivative of tensors at $x$ will be called $\nabla$.
It involves only Christoffel symbols with indices $\mu$, $\nu$, $\rho$ etc.
Tensors evaluated at $\varphi(x)$ have to be treated as scalars 
(they are inert under changes of frame at $x$).
Thus for example $\nabla_\nu g_{\mu\nu}=0$ as usual, but
\be
\nabla_\nu g_{\beta\alpha}(\varphi)
=\partial_\nu\varphi^\gamma\partial_\gamma g_{\beta\alpha}(\varphi)\ .
\label{ennio}
\ee
There is also a notion of covariant derivative on tensors at $\varphi$.
We will write $D$ for this type of covariant derivative.
For example
\be
D_\mu\omega_\alpha(\varphi)=
\partial_\mu\varphi^\gamma
D_\gamma\omega_\alpha\ ;
\quad
D_\gamma\omega_\alpha=
\partial_\gamma\omega_\alpha
- \Gamma_\gamma{}^\beta{}_\alpha \omega_\beta(\varphi)\ .
\label{ettore}
\ee
Note that in spite of carrying an index, $\varphi^\alpha$ are not
vectors. Their covariant derivative is the same as the ordinary
partial derivative. 
Thus we will use interchangeably the notation
$\partial_\mu\varphi^\alpha$ and $\nabla_\mu\varphi^\alpha$,
but when the derivative index is raised we always use $\nabla$, 
so
$g^{\rho\mu}\partial_\mu\varphi^\alpha=\nabla^\rho\varphi^\alpha$.

We shall also encounter objects that have indices of both types.
For example, $\nabla_\mu\varphi^\alpha$ is a covariant vector
at $x$ and a contravariant vector at $\varphi(x)$.
\footnote{It is a section of $T^*M\otimes\varphi^*TM$.
The connection in this bundle is the tensor product of the
Levi-Civita connection in $T^*M$ and the 
pullback of the Levi-Civita connection in $TM$.}
In this case we find it convenient to use a covariant derivative
$\nabla$ that covariantizes only the index $\mu$, while $\cD$
is the full covariant derivative that covariantizes both indices:
\bea
\nabla_\rho \nabla_\mu\varphi^\alpha
&=&\partial_\rho \nabla_\mu\varphi^\alpha
-\Gamma_\rho{}^\sigma{}_\mu\nabla_\sigma\varphi^\alpha\ .
\label{defD}
\\
\cD_\rho \nabla_\mu\varphi^\alpha
&=&\nabla_\rho \nabla_\mu\varphi^\alpha
+\partial_\rho\varphi^\gamma
\Gamma_\gamma{}^\alpha{}_\beta\nabla_\mu\varphi^\beta\ ,
\label{defnabla}
\eea

Let $g_{\mu\nu}$ be a metric on $M$.
The basic dynamical variable $\varphi$ is a diffeomorphism
of $M$ to itself.
In analogy to (\ref{B1}), the basic building block of the action
is the tensor
\be
B^{\mu}{}_\nu = g^{\mu\rho}(\varphi^* g)_{\rho\nu}
= g^{\mu \rho}\partial_\rho \varphi^\beta g_{\beta \alpha}(\varphi)\partial_\nu \varphi^\alpha\ .
\label{defB}
\ee
In the r.h.s. we have written the factors in a particular order 
that calls for the use of matrix notation.
If we denote $J_\mu{}^\alpha=\partial_\mu\varphi^\alpha$
the Jacobian of $\varphi$, we can write (\ref{defB}) in the form
$$
B = g^{-1}\,(\varphi^* g)
= g^{-1}Jg(\vp)J^T\ .
$$
This notation is often convenient in calculations.

Let us now construct scalars out of this tensor. 
Since $B$ is assumed to transform as a mixed tensor,
the invariants are traces of products of powers of $B$.
Only four such terms are algebraically independent. 
An arbitrary scalar can be expressed, generally in a nonlinear way, 
via four chosen independent scalars. 
The natural choice of four independent scalars is 
\be
\tau_k = \Tr(B^k)\ ,\qquad k = 1,2,3,4\ .
\ee
Our action is then 
\be
S=\int dx\,\sqrt{g}\,\cL(\tau_1,\tau_2,\tau_3,\tau_4)\ ,
\label{actgh}
\ee
where $\cL$ is the Lagrangian, an arbitrary function of the traces.
In the following we will sometimes consider the simplest case 
\be
\cL=-\frac{1}{2}f^2\tau_1
\label{lag1}
\ee
where $f$ is a coupling. In mathematical literature (where $g$ is a Riemannian metric) the stationary points are the harmonic maps 
\cite{eellslemaire}.
Another particularly interesting Lagrangian is 
\be
\mathcal{L}=c(\tau_1^2-\tau_2)\ .
\label{lag2}
\ee
We shall discuss some of its properties in Sect. 3.2.

Let us now discuss the invariances of the theory.
As long as $\varphi$ are the only dynamical fields,
and $g$ is a fixed metric, the only symmetries of the action
are the isometries of $g$.
Thus, generically, there are no symmetries.
Even though we will not discuss dynamical gravity in this paper,
we shall consider the invariances of the action 
when the metric is allowed to transform.
This becomes relevant when one couples the fields
to a dynamical metric, and will also be important later in the
discussion of the energy-momentum tensor and its conservation.

A diffeomorphism $\psi$ acts on tensors on $M$ in the standard way,
in particular the action on covariant tensors is by pullback.
For example, the metric $g$ is transformed to $g'=\psi^* g$.
The action on $\varphi$ can be either
by left or by right composition.
Under right composition
\be
\varphi\mapsto\varphi'=\varphi\circ\psi
\ee
we have
\be
\varphi^* g
\mapsto
\varphi^{\prime*}g'
=\psi^*\varphi^*\psi^* g\ ,
\ee
which is not the correct transformation of a covariant tensor.
Thus $B$ does not transform properly as a mixed tensor.
Similarly under left composition
\be
\varphi\mapsto\varphi'=\psi^{-1}\circ\varphi\ ,
\ee
we find that the pullback of $g$ is invariant:
\be
\varphi^* g
\mapsto
\varphi^{\prime*}g'
=\varphi^*\psi^{-1*}\psi^* g
=\varphi^* g\ .
\ee
Thus again $B$ does not transform properly.
This was to be expected, because
right composition (pullback) is the natural transformation for
maps {\it on} spacetime (i.e. maps having spacetime
as domain)
and left composition is the natural transformation for
maps {\it into} spacetime (i.e. having spacetime as target).
Since a diffeomorphism $\varphi$ is simultaneously
a map on and in spacetime, one should act both ways.
Indeed, consider now the ``diagonal subgroup''
acting by conjugation
\be
\varphi\mapsto\varphi'=\psi^{-1}\circ\varphi\circ\psi\ .
\ee
In this case we find that the pullback of the metric
transforms as a covariant tensor:
\be
\varphi^* g
\mapsto
\varphi^{\prime*}g'
=\psi^*\varphi^*\psi^{-1*}\psi^* g
=\psi^*\varphi^* g\ .
\ee
This leads to the correct transformation of $B$.
We have established that the action $S(\varphi)$ is invariant
under $\diffM$, acting on the metric by pullback
and on $\varphi$ by conjugation.

It is important to appreciate the following point:
whereas the actions of $\diffM$ on itself by left and right
composition are transitive (every $\varphi\in \diffM$ can be
mapped to any other $\varphi'$ by a right- or left-composition)
and free (there are no fixed points),
the diagonal action is not.
In fact, the diagonal action leaves the identity map invariant.

In Appendix B we discuss closely related models
where the domain and target spaces are kept separate,
and show that the action is separately invariant under left- and right-diffeomorphisms.
In contrast to those models,
neither of these group actions leaves the action
(\ref{actgh}) invariant.

Finally we observe that the identification of the spacetime 
and target space metrics
leads to peculiar properties also from the point of view
of dimensional analysis.
Since the two metrics appearing in $B$ are functions of 
$x$ and $\varphi(x)$, respectively,
it is most natural to assume that the fields $\varphi^\alpha$
have the same dimension as the coordinates $x^\mu$.
\footnote{It is generally the case that when the
Lagrangian contains non-polynomial interactions,
the fields should be dimensionless.
This, together with $[\varphi]=[x]$
leads to dimensionless coordinates,
a choice that we find most natural also for other reasons.
However, we do not need to commit to this choice here.}
We will assume this throughout this paper.
Then, the tensor $B^\mu{}_\nu$ is dimensionless,
and so are the traces $\tau_n$.
The couplings $f^2$ and $c$ in the Lagrangians (\ref{lag1})
and (\ref{lag2}) have mass dimension equal to the spacetime dimension $d$.
\footnote{Note that in ordinary nonlinear sigma models $f^2$ would have
dimension $d-2$ and $c$ would have dimension $d-4$.
The difference is due to the identification $h=g$.
Also note that one could absorb $f^2$ in the
metric, theerby making it dimensionless.}
In fact, the coefficient of any monomial in the $\tau_n$
must have dimension $d$.
Since the Lagrangian $\cL$ must have dimension $d$,
we could extract an overall factor $f^2$ with dimension $d$
and write $\cL=f^2\tilde \cL$,
where $\tilde\cL$ is a purely numerical function.
As usual, one expects that $\tilde\cL$ does not
contain exceedingly large or exceedingly small coefficients.
Then, $f^2$ is the only characteristic scale of the theory
and it will be related to the scale of the cosmological constant.

\subsection{Equation of motion of $\varphi$} 
\label{general}

Let $\varphi(t)$ be a one-parameter family of maps $M\to M$.
The derivative 
\be
\frac{d\varphi^\alpha(t)}{dt}\Bigg|_{t=0}=v^\alpha
\label{sectv}
\ee
is a section of the vector bundle $\varphi^*TM$.
The equation of motion is obtained by setting to zero the
directional derivative of the action (\ref{actgh})
along an arbitrary vector $v$:
\bea
0=vS\!\!&=&\!\!\frac{dS(\varphi(t))}{dt}\Bigg|_{t=0}
\nonumber\\
&=&\!\!
\int dx \sqrt{g} \sum_n n\frac{\pa \cL}{\pa \tau_n} \Tr B^{n-1}
\frac{\partial B}{\partial t}\Bigg|_{t=0}
\nonumber\\
&=&\!\!
\int dx \sqrt{g} \sum_n n\frac{\pa \cL}{\pa \tau_n} (B^{n-1})^{\sigma\nu} 
\left(2\pa_\nu v^\beta g_{\beta \alpha}(\varphi)
+\pa_\nu \varphi^\beta v^\gamma \pa_\gamma g_{\beta \alpha}(\varphi)\right)\pa_\sigma \varphi^\alpha\ .
\eea
We used the fact that $(B^{n-1})^{\sigma\nu}$ is symmetric.
In the first term we have to extract the variation $v$
from the derivative.
We use that 
$\partial_\nu v^\beta=\nabla_\nu v^\beta$
and use the standard rules for integration by parts.
In this way we find
\bea
0=\,v S\!&=&\!
\int dx \sqrt{g}\,v^\alpha \Bigg\{
-\nabla_\mu \left [2 \sum_n n\frac{\pa \cL}{\pa \tau_n} (B^{n-1})^\mu{}_\rho g^{\rho \tau} \pa_\tau \varphi^\beta g_{\beta \alpha}(\varphi) \right] \nn 
&& 
\qquad\qquad\qquad
+ \sum_n n\frac{\pa \cL}{\pa \tau_n}(B^{n-1})^\mu{}_\rho g^{\rho \tau} \pa_\tau \varphi^\beta \pa_\alpha g_{\beta \gamma}(\varphi)\pa_\mu \varphi^\gamma \Bigg\}
\nn
\!&=&\!
-2\int dx \sqrt{g}\,v^\alpha g_{\alpha\beta}(\varphi)
\cD_\mu \left[\sum_n n\frac{\pa \cL}{\pa \tau_n} (B^{n-1})^{\mu\nu}\pa_\nu \varphi^\beta  \right]\ .
\label{eom1}
\eea
In the last step, we used (\ref{ennio}) in the terms
where $\nabla_\mu$ hits $g_{\beta\alpha}(\varphi)$.
The resulting terms combine with the second line to
produce a Christoffel symbol that enters in
the covariant derivative $\cD_\mu$.
Thus the equation of motion reads
\be
\cD_\mu \left[\sum_n n\frac{\pa \cL}{\pa \tau_n} (B^{n-1})^{\mu\nu}\pa_\nu \varphi^\beta  \right]=0\ .
\label{eom}
\ee

One can be more explicit in special cases.
For example, in the case of the action (\ref{harmonic}),
the only term in the sum has $\frac{\pa \cL}{\pa\tau_1}=1$
and $B^0=\mathbf{1}$, or
$(B^0)^{\mu\nu}=g^{\mu\nu}$.
Using the aforementioned rules, and recalling that
$\partial_\mu\varphi^\alpha=\nabla_\mu\varphi^\alpha$,
this leads to the equation for harmonic maps:
\begin{equation}
\cD^\mu \nabla_\mu \varphi^\beta = 0\ .
\label{eomharm}
\end{equation}

Applying this to the other special case (\ref{lag2}), the equation of motion becomes
\begin{equation}
\cD_\mu [\tau_1\nabla^{\mu}\varphi^\beta-B^{\mu \nu}\pa_\nu\varphi^\beta] =0\ .
\end{equation}

It will be useful to rewrite (\ref{eomharm}) in another way.
Expanding the covariant derivatives one has
\be
0=g^{\mu\nu}
\left(\pa_\mu\pa_\nu\varphi^\alpha
-\Gamma_\mu{}^\rho{}_\nu\pa_\rho\varphi^\alpha
+\pa_\mu\varphi^\beta \Gamma_\beta{}^\alpha{}_\gamma
\pa_\nu\varphi^\gamma\right)\ .
\label{eomharm2}
\ee
Since $\varphi$ is a diffeomorphism,
from the transformation properties of the connection
we can rewrite the first two terms as
\be
\pa_\mu J_\nu{}^\alpha
-\Gamma_\mu{}^\rho{}_\nu J_\rho{}^\alpha
=-J_\mu{}^\beta J_\nu{}^\gamma \Gamma'_\beta{}^\alpha{}_\gamma
\ee
and therefore the equation for harmonic diffeomorphisms
amounts to the statement that
\be
0=g^{\prime\beta\gamma}
\left(\Gamma_\beta{}^\alpha{}_\gamma
-\Gamma'_\beta{}^\alpha{}_\gamma\right)\ ,
\label{alteomharm}
\ee
where $g^{\prime\beta\gamma}=g^{\mu\nu}\pa_\mu\varphi^\beta \pa_\nu\varphi^\gamma$ and $\Gamma'$ are the Christoffel symbols of $g'$.
Since the difference of two connections is a tensor,
this is a covariant statement.

\subsection{Energy-momentum tensor}

Next, we vary the action with respect of $g_{\mu \nu}(x)$:
\bea
\delta S &=& \int dx \sqrt{g}\,\Bigg\{\frac{1}{2} 
g^{\mu\nu} \delta g_{\mu\nu} \cL 
+ \sum_n n\frac{\pa \cL}{\pa \tau_n} \Tr B^{n-1}\delta B \Bigg\}\nn
&=&
\int dx \sqrt{g}\,\Bigg\{
\frac{1}{2} g^{\mu\nu} \delta g_{\mu\nu} \cL 
+ \sum_n n\frac{\pa \cL}{\pa \tau_n} (B^{n-1})^\sigma{}_\mu 
\bigg(-g^{\mu\rho} \delta g_{\rho\tau} g^{\tau\lambda} \pa_\lambda \vp^\gamma g_{\gamma \delta}(\vp)\pa_\sigma \vp^\delta \nn
&& \qquad \qquad\qquad\qquad\qquad\qquad\qquad\qquad\qquad\qquad
+ g^{\mu\lambda} \pa_\lambda \vp^\alpha \delta g_{\alpha \beta}(\vp)\pa_\sigma \vp^\beta \bigg)  \Bigg\} \ .
\nonumber
\eea
We can obtain the energy-momentum tensor by straightforwardly
using the rules of variational calculus and evaluating
\be
T^{\mu\nu}=
\frac{2}{\sqrt{g}}
\frac{\delta S}{\delta g_{\mu\nu}}\ .
\label{enmom}
\ee
It is perhaps more instructive to observe 
that in the last term the variation appears in the combination
$\pa_\lambda\vp^\alpha\pa_\sigma \vp^\beta
\delta g_{\alpha\beta}(\vp)
=(\varphi^*\delta g)_{\mu\nu}(x)$.
We exploit the fact that the integral does not change if we
replace the integrand by its transform under a diffeomorphism.
Let $x^{\prime\alpha}=\varphi^\alpha(x)$,
and denote by a prime all trasformed tensors.
Covariant tensors are pulled back by $\varphi^{-1}$ and
contravariant tensors are pushed forward with $\varphi$:
$A'=\varphi_* A$, $g'=\varphi^{-1*}g$, and
$(\varphi^*\delta g)'=\delta g$.
Then the last term can be manipulated as follows:
$$
\int dx\sqrt{g(x)}\,A^{\mu\nu}(x)(\varphi^*\delta g)_{\mu\nu}(x)
=\int dx'\sqrt{g'(x')}\,A^{\prime\alpha\beta}(x')
\delta g_{\alpha\beta}(x')\ ,
$$
where $x'$ is to be regarded as independent integration variable.
Then, from the definition given above, we obtain
\be
T^{\mu\nu}\!=\!
g^{\mu\nu}\cL
-2\sum_n n\frac{\pa \cL}{\pa \tau_n} (B^{n-1})^{\sigma\mu}  \nabla^\nu\vp^\gamma g_{\gamma \delta}(\vp)\nabla_\sigma \vp^\delta  +2\frac{\sqrt{g'}}{\sqrt{g}} \sum_n n\left( \frac{\pa \cL}{\pa \tau_n} \right)' (B^{\prime\,n-1})^{\mu\nu}\ .
\label{emtensor}
\ee
\bigskip
In particular for the Lagrangian (\ref{lag1})
\begin{equation}
T^{\mu\nu}=f^2\left[
\nabla^\mu\varphi^\alpha\nabla^\nu\varphi^\beta g_{\alpha\beta}(\varphi)
-\frac{1}{2}g^{\mu\nu}
g^{\rho\sigma}
\nabla_\rho\varphi^\alpha\nabla_\sigma\varphi^\beta g_{\alpha\beta}(\varphi)
-\frac{\sqrt{g'}}{\sqrt{g}}g^{\prime\mu\nu}\right]\ ,
\label{enmomharm}
\end{equation}
\bigskip
whereas for (\ref{lag2})
\be
T^{\mu\nu}=c\,g^{\mu\nu}(\tau_1^2-\tau_2) 
- 4c(\tau_1 g^{\sigma\mu}-B^{\sigma\mu})\nabla^\nu\varphi^\gamma g_{\gamma \delta}(\varphi)\nabla_\sigma\varphi^\delta
+4\frac{\sqrt{g'}}{\sqrt{g}} 
c(\tau'_1 g^{\prime\mu\nu}-B^{\prime\mu\nu})\ .
\ee

\subsection{Diffeomorphism invariance, EOM and EM conservation}

Let us begin by recalling the general argument relating diffeomorphism
invariance to EM conservation.
Given an action $S(\varphi,g)$
for the ``matter'' fields $\varphi$ coupled to a metric $g$,
its variation under an infinitesimal diffeomorphism $\xi$ is
\be
\delta_\xi S=
\int dx\left[\frac{\delta S}{\delta\varphi^\alpha}
\delta_\xi\varphi^\alpha
+\frac{\delta S}{\delta g_{\mu\nu}}\delta_\xi g_{\mu\nu}\right]\ .
\label{vardiffS}
\ee
We define the equation of motion 
\be
E^\alpha=\frac{1}{\sqrt g}\frac{\delta S}{\delta\varphi^\alpha}\ .
\ee
An infinitesimal diffeomorphism is defined by
\be
\delta_\xi x^\mu=-\xi^\mu(x)\ ,
\ee
where $\xi$ is a vector field.
The infinitesimal variation of $\varphi$ is
\be
\delta_\xi\varphi^\alpha(x)
=\xi^\lambda\partial_\lambda\varphi^\alpha-\xi^\alpha(\varphi)\ ,
\ee
where the first term comes from the
right composition and the second from the left composition.
The variation of any tensor $T$ is its Lie derivative
$\delta_\xi T={\cal L}_\xi T$.
For the metric
$$
\delta_\xi g_{\mu\nu}
=\xi^\rho\partial_\rho g_{\mu\nu}
+g_{\mu\rho}\partial_\nu\xi^\rho
+g_{\rho\nu}\partial_\mu\xi^\rho
=\nabla_\mu\xi_\nu+\nabla_\nu\xi_\mu\ .
$$

Inserting these formulae and (\ref{enmom}) in (\ref{vardiffS}),
invariance of the action implies
\be
0=\int d^4x\sqrt{g}
\left[(\xi^\tau\nabla_\tau\vp^\alpha - \xi^\alpha(\vp))E_\alpha
-\xi_\mu\nabla_\nu T^{\mu\nu}
\right]\ .
\label{diffinv}
\ee
We find, as expected, that $E_\alpha=0$ implies EM conservation.

In the case of our theory of diffeomorphisms, there is more to
be learned.
We observe that for a generic $\varphi$,
the coefficient of $E_\alpha$ is non-vanishing
and therefore, conversely EM conservation also 
generically implies the EOM.
This makes sense, because both statements amount
to four second order differential equations for the fields.
However, for the identity map $\varphi^\alpha=x^\alpha$,
the coefficient of $E_\alpha$ vanishes and therefore
this implication does not hold.
This is a consequence of the identity map being
a fixed point of the action of the diffeomorphism group.

One could further explicitly compute the divergence of the EM tensor.
This calculation is very complicated in general,
but we can do it in the case of the Lagrangian (\ref{lag1}).
It turns out to be useful to split the EM tensor in two parts:
\be
T^{\mu\nu}=T_{(R)}^{\mu\nu}+T_{(L)}^{\mu\nu}\ ,
\label{emsplit}
\ee
where the first part arises from the variations with respect
to $g_{\mu\nu}$ and consists of the first two terms in (\ref{enmomharm}),
the second part comes from variation with respect to
$g_{\alpha\beta}(\varphi)$ and consists of the third term
in (\ref{enmomharm}).
Interestingly, these two parts are separately conserved.
The conservation of $T_{(R)}^{\mu\nu}$ works exactly as
for a nonlinear sigma model:
$$
\nabla_\mu T_{(R)}^{\mu\nu}=
\nabla^2\varphi^\alpha\nabla^\nu\varphi^\beta 
g_{\alpha\beta}(\varphi)
+\left(\nabla^\mu\varphi^\alpha\nabla^\nu\varphi^\beta
\nabla_\mu\varphi^\gamma
-\frac12\nabla^\mu\varphi^\alpha\nabla_\mu\varphi^\beta
\nabla_\nu\varphi^\gamma
\right)
\partial_\gamma g_{\alpha\beta}(\varphi)\ .
$$
The two terms in parentheses reconstruct a Christoffel symbol,
and the whole expression is then seen to be proportional
to the EOM, written in the form (\ref{eomharm2}).

For the second part,
\bea
\nabla_\mu T_{(L)}^{\mu\nu}&=&
-f^2\frac{1}{\sqrt{g}}
\nabla_\mu(\sqrt{g'}g^{\prime\mu\nu})
\nonumber\\
&=&-f^2\frac{1}{\sqrt{g}}
\left(
\partial_\mu(\sqrt{g'}g^{\prime\mu\nu})
+\sqrt{g'}\,\Gamma_\mu{}^\nu{}_\rho g^{\prime\mu\rho}
\right)
\nonumber\\
&=&-f^2\frac{\sqrt{g'}}{\sqrt{g}}\,g^{\mu\rho}
\left(\Gamma_\mu{}^\nu{}_\rho
-\Gamma'_\mu{}^\nu{}_\rho\right)
\eea
which vanishes due to the EOM
written in the form (\ref{alteomharm}).

While these statements are easy to check in the case
of the Lagrangian (\ref{lag1}), calculating the divergence of
the EM tensor in the general case is very complicated.
Still, the preceding statements remain true.
One can see this by following in detail the proof of
invariance of the action under infinitesimal diffeomorphisms,
which is given in Appendix A.
One can see there that the terms coming from variations of $g_{\mu\nu}$
(i.e. the divergence of $T_{(R)}^{\mu\nu}$)
and those coming from variations of $g_{\alpha\beta}(\varphi)$
(i.e. the divergence of $T_{(L)}^{\mu\nu}$)
cancel separately against terms coming from the variation
of $\varphi$ (i.e. the EOM).
More precisely, the differential identity (\ref{diffinv})
can be seen as the sum of two separate identities
\bea
\int d^4x\sqrt{g}
\left[\xi^\tau\nabla_\tau\vp^\alpha E_\alpha
-\xi_\mu\nabla_\nu T_{(R)}^{\mu\nu}
\right]&=&0\ ,
\\
\int d^4x\sqrt{g}
\left[- \xi^\alpha(\vp))E_\alpha
-\xi_\mu\nabla_\nu T_{(L)}^{\mu\nu}
\right]&=&0\ .
\label{diffinv2}
\eea

This is surprising, because the invariance group of the action
has four parameters but it seems to imply eight
differential identities.
This can be explained by looking at the models as special
cases of theories with different domain and target space.
In these cases, as explained in Appendix B,
the invariance group consists separately of left and
right diffeomorphisms and therefore implies eight
differential identities.
The models with identical domain and target space
are obtained by choosing a preferred diffeomorphism,
and this does not invalidate the identities.

\section{The identity solution}

The identity $\varphi=Id_M$ is represented, 
in any local coordinate system, by
\be
\varphi^\alpha=x^\alpha\ .
\ee
In this case the Jacobian reduces to 
$\partial_\mu\varphi^\alpha=\delta^\alpha_\mu$,
$B^{\mu\nu}=g^{\mu\nu}$
and all the traces become constant: $\tau_n=4$ for $n=1,2,3,4$.
Thus $\cL$ and its derivatives are just constants
and the EOM (\ref{eom}) are satisfied.

The energy-momentum tensor of the solution
becomes proportional to the metric.
The last two terms in (\ref{enmom}) cancel out and
\be
T^{\mu\nu}=g^{\mu\nu}\cL
-2\sum_n n\frac{\pa \cL}{\pa \tau_n} g^{\mu\nu} 
+2\sum_n n\Big( \frac{\pa \cL}{\pa \tau_n} \Big) g^{\mu\nu} 
= g^{\mu\nu}\cL(4,4,4,4)\ .
\ee
We can therefore interpret this energy-momentum tensor
as an effective cosmological constant
$\Lambda=8\pi G\, \cL(4,4,4,4)$.
As already anticipated, if we assume that 
$\tilde \cL(4,4,4,4)$ is a number of order one,
the effective cosmological constant is $\Lambda\approx 8\pi G f^2$.

We will now discuss the stability of the identity solution.
This can have two possible meanings: in Euclidean signature one
asks whether the action has a minimum at the solution;
in Lorentzian signature one asks whether the energy
has a minimum.
We begin by discussing the simpler problem of Euclidean stability.
Since the Euclidean action is also identical to the energy
of a static field configuration (in one dimension more)
this analysis also says something about the stability
of static configurations under static deformations.
Full Lorentzian stability will be briefly discussed in Section 3.3,
where we shall refer to existing results in the literature.

\subsection{The second variation}

Here we compute the Hessian of the action at the identity solution.
This is needed to establish whether a Euclidean solution is stable,
and is also needed in the study of linearized perturbations.
Let $\varphi(t,s)$ be a two-parameter family of maps.
We take the double derivatives at $t=s=0$.
We let
\be
\frac{d\varphi^\alpha(t,s)}{dt}\Bigg|_{t=s=0}=v^\alpha\ ,
\qquad
\frac{d\varphi^\alpha(t,s)}{ds}\Bigg|_{t=s=0}=w^\alpha\ .
\ee
The Hessian is defined by
\bea
H(v,w)&=&\frac{\partial^2 S(\varphi(t,s))}{\partial t\partial s}\Bigg|_{t=s=0}
\nonumber\\
&=&-2\frac{\partial}{\partial s}
\int dx\sqrt{g}\,\frac{\partial\varphi^\alpha}{\partial t} 
g_{\alpha\beta}\,
\left(\cD_\mu\sum_n n\frac{\partial \cL}{\partial\tau_n}
(B^{n-1})^{\mu\nu}
\partial_\nu\varphi^\beta\right)\Bigg|_{t=s=0}\ .
\eea
The derivatives with respect to $s$ of the terms
$\frac{\partial\varphi^\alpha}{\partial t}$ and $g_{\alpha \beta}$
are proportional to the EOM,
and since we are interested in the variation around a solution,
we can neglect them.
Acting with the $s$-derivative on all the remaining occurrences of
$\varphi$, and evaluating at $t=s=0$,
which correspond to $\varphi=Id_M$,
the round bracket becomes
\bea
&&\sum_n n\frac{\partial \cL}{\partial\tau_n}
\cD^\mu\partial_\mu w^\beta
+\sum_n n\frac{\partial \cL}{\partial\tau_n}
\cD_\mu\left(\frac{\partial}{\partial s}(B^{n-1})^{\mu\beta}
\Bigg|_{s=0}\right)
+\sum_{n,m} n\frac{\partial^2 \cL}{\partial\tau_n\partial\tau_m}
D^\beta\left(\frac{\partial\tau_m}{\partial s}\Bigg|_{s=0}\right)
\nonumber\\
&&
+\sum_n n\frac{\partial \cL}{\partial\tau_n}
g^{\mu\delta}
\left(\partial_\mu w^\gamma\Gamma_\gamma{}^\beta{}_\delta
+w^\epsilon\partial_\epsilon\Gamma_\mu{}^\beta{}_\delta\right)\ .
\label{enrico}
\eea
Now consider the general formula
\bea
\cD_\mu D_\nu w^\delta&=&
\nabla_\mu\nabla_\nu w^\delta
+\nabla_\mu\nabla_\nu\varphi^\gamma \Gamma_\gamma{}^\delta{}_\beta w^\beta
+\partial_\nu\varphi^\gamma\partial_\mu\varphi^\beta
\partial_\beta\Gamma_\gamma{}^\delta{}_\epsilon w^\epsilon
\nonumber\\
&&
+\partial_\nu\varphi^\gamma\Gamma_\gamma{}^\delta{}_\beta\partial_\mu w^\beta
+\partial_\mu\varphi^\gamma\Gamma_\gamma{}^\delta{}_\epsilon
\left(\partial_\nu w^\epsilon
+\partial_\nu\varphi^\zeta\Gamma_\zeta{}^\epsilon{}_\beta w^\beta
\right)
\eea
and specialize to the identity map.
Comparing this with the preceding formula, 
and using $\nabla_\mu\nabla_\nu\varphi^\beta=-\Gamma_\mu{}^\beta{}_\nu$,
we see that the first and the last term in (\ref{enrico})
combine to give
$$
\sum_n n\frac{\partial \cL}{\partial\tau_n}
\left(\cD^\mu D_\mu w^\beta+R^\beta{}_\gamma w^\gamma\right)\ .
$$
At the identity, we can convert all indices
to $\mu$, $\nu$ etc and $\cD^\mu D_\mu=\nabla^2$.
In the remaining two terms we have
$$
\frac{\partial}{\partial s}(B^{n-1})^{\mu\nu}
\Bigg|_{s=0}
=(n-1)\left(\nabla^\mu w^\nu+\nabla^\nu w^\mu\right)
$$
and
$$
\frac{\partial\tau_m}{\partial s}\Bigg|_{s=0}
=2m \nabla_\mu w^\mu\ .
$$
Putting everything together, the second variation of the action
around the identity is
\bea
H(v,w)&=&2
\int dx\sqrt{g}\Bigg[
\sum_n n^2\frac{\partial \cL}{\partial\tau_n}
v^\mu\left(-\nabla^2g_{\mu\nu}-R_{\mu\nu}\right)w^\nu
\nonumber\\
&&
+\left(\sum_n n(n-1)\frac{\partial \cL}{\partial\tau_n}
+2\sum_{m,n} mn\frac{\partial^2 \cL}{\partial\tau_n\partial\tau_m}\right)
(\nabla_\mu v^\mu)(\nabla_\nu w^\nu)
\Bigg]\ .
\eea
We can write $H(v,w)=\langle v,L w\rangle$,
where $\langle\,,\rangle$ is the natural inner product in the space
of sections of $\varphi^*TN$ and $L$ represents the differential operator
\bea
L&=&
\sum_n n^2\frac{\partial \cL}{\partial\tau_n}
\left(-\nabla^2g_{\mu\nu}-R_{\mu\nu}\right)
\nonumber\\
&&
-\left(\sum_n n(n-1)\frac{\partial \cL}{\partial\tau_n}
+2\sum_{m,n} mn\frac{\partial^2 \cL}{\partial\tau_n\partial\tau_m}\right)
\nabla_\mu\nabla_\nu \ .
\eea
Note that the factors involving $\cL$ are constants.
Using the rules for integrations by parts, one sees that the Hessian is symmetric:
\be
H(v,w)=H(w,v)
\ee
which also means that the differential operator $L$
is self-adjoint.
\bigskip

Clearly, stability hinges on the form of the Lagrangian.
As a first example consider the action
for harmonic diffeomorphisms (\ref{harmonic}). 
Its Hessian is \cite{mazet,smith}:
\bea
H(v, w) = 2 \int dx \sqrt{g} \,
v^\mu\,\left(-\nabla^2g_{\mu\nu}-R_{\mu\nu}\right)w^\nu\ .
\eea
Hence we have to study te spectrum of the Laplace-type operator
\bea
L_1 = -g_{\mu\nu}D^2-R_{\mu\nu}\ .
\eea
This spectrum is known for spheres.
It consists of transverse and longitudinal fields.
The lowest transverse eigenfunctions are
the $d(d+1)/2$ Killing vectors,
which are zero modes.
This is related to the fact that every isometry is harmonic.
The lowest longitudinal eigenfunctions are
\be
w_\nu^i = \pa_{\nu}\phi^i\ ,
\label{negative}
\ee
where $\phi^i$ are cartesian coordinates of the flat
Euclidean space in which the sphere is embedded.
These are conformal Killing vectors and
have eigenvalue $-\frac{d-2}{d(d-1)}R$
and multiplicity $d+1$.
In $d=2$ they are zero modes; this is related
to the fact that in $d=2$ conformal isometries are harmonic 
\cite{chowlu}.
All other eigenvalues are positive.
Thus, the identity is unstable as a harmonic map
of spheres in $d>2$; in $d=2$ it is stable and belongs to
a six-parameter family of degenerate solutions.
\bigskip

As another example we consider the model (\ref{lag2}).
In this case the Hessian is
\bea
H(v,w)&=&4
\int dx\sqrt{g}\Bigg[
(d-2)
v^\mu\left(-\nabla^2g_{\mu\nu}-R_{\mu\nu}\right)w^\nu
+(\nabla_\mu v^\mu)(\nabla_\nu w^\nu)
\Bigg]
\eea
and, integrating the second term by parts, the associated operator reads
\bea
L_2 =(-\nabla^2g_{\mu\nu}-R_{\mu\nu})
-\frac{1}{d-2}\nabla_\mu\nabla_\nu\ .
\label{Lnoghost}
\eea
Clearly the spectrum on transverse vectors is the same as 
that of $L_1$, but it may differ on longitudinal vectors,
and the new contribution is positive,
so that the identity may become stable.
In fact, the additional non-minimal term, acting on
the eigenfunctions (\ref{negative}), gives
$R/(d-1)(d-2)$, so that the eigenvalue becomes
$-\frac{d-4}{d(d-2)}R$.
It is negative in all dimensions except four, were it is zero.
Also in this case, the zero modes are related to
the infinitesimal isometries and conformal isometries.

\subsection{Global Euclidean bounds}

The second variation of the action gives information
about the local stability of a solution,
but there are some cases where absolute bounds
on the action can be derived.
In this section we assume that $M$ is compact without boundary.
We use the totally antisymmetric tensor
$\eta_{\mu_1\ldots\mu_d}=\sqrt{g}\,\varepsilon_{\mu_1\ldots\mu_d}$
where $\varepsilon$ is the tensor density with components $\pm1$.
It is the volume form on $M$, such that 
$V=\int_M\eta=\int d^dx\sqrt{g}$ is the volume.
The winding number is
\be
W=\frac{1}{d!V}\int d^dx\sqrt{g}\, \eta^{\mu_1\ldots\mu_d}
J_{\mu_1}{}^{\alpha_1}\ldots
J_{\mu_d}{}^{\alpha_d}
\eta_{\alpha_1\ldots\alpha_d}
\ee
and is equal to one for orientation-preserving diffeomorphisms.

The action for harmonic maps in two Euclidean dimensions is
$S=\frac12 f^2\int d^2x\sqrt{g}\,\tau_1$.
We define the double dual $^*\!J^*$ by
$^*\!J^*_{\mu}{}^{\alpha}
=\eta_{\mu}{}^{\rho}
J_{\rho}{}^{\gamma}\eta_{\gamma}{}^{\alpha}$.
Integrating the square of $J\pm^*\!J^*$ one
obtains the well-known bound  $S\geq V f^2 |W|$, \cite{belapol}
and since $|W|=1$ for diffeomorphisms,
\be
S\geq V f^2\ .
\ee
The absolute minima are the maps for which $J=\pm^*\!J^*$. 
Indeed, the identity solves this equation.

There is a parallel example in four Euclidean dimensions.
This time we consider the Lagrangian (\ref{lag2}).
Defining the antisymmetric tensor
$K_{\mu\nu}{}^{\alpha\beta}=J_{[\mu}{}^{[\alpha}
J_{\nu]}{}^{\beta]}$,
and the inner product
$$
(K,K')=
\frac12 \int d^4x\sqrt{g}\,
g^{\mu\rho}g^{\nu\sigma}
K_{\mu\nu}{}^{\alpha\beta}K'_{\rho\sigma}{}^{\gamma\delta}
g_{\alpha\gamma}g_{\beta\delta}\ ,
$$
this action can be rewritten
\be
S=
4c||K||^2\equiv
2 c\int d^4x\sqrt{g}\,
g^{\mu\rho}g^{\nu\sigma}
K_{\mu\nu}{}^{\alpha\beta}K_{\rho\sigma}{}^{\gamma\delta}
g_{\alpha\gamma}g_{\beta\delta}\ .
\ee
We can define $^*\!K^*$, the double dual of $K$, by
\be
^*\!K^*_{\mu\nu}{}^{\alpha\beta}
=\frac14\eta_{\mu\nu}{}^{\rho\sigma}
\eta^{\alpha\beta}{}_{\gamma\delta}
K_{\rho\sigma}{}^{\gamma\delta}\ .
\ee
Since $||^*\!K^*||^2=||K||^2$,
we have
\be
0\leq||K\pm^*\!\!K^*||^2
=2||K||^2\pm2(K,^*\!\!K^*)
=\frac{1}{2c}S\pm 6 VW[\varphi]\ ,
\ee
where $W$ is the winding number of $\varphi$.
Since diffeomorphisms have $|W|=1$, we have the absolute bound
\be
S\geq 12\, cV \ .
\ee
The bound is saturated by maps for which $H$ is double-self-dual,
and the identity has this property.

For generic metric, the identity will be an isolated solution.
In the presence of isometries and/or conformal isometries,
it will be an element of continuous degenerate families of solutions,
as we have seen in the end of the preceding section.

\subsection{Lorentzian stability}

As long as we restrict ourselves to dynamical diffeomorphisms
in a fixed external metric, 
there is no difference, regarding the issues of Lorentzian stability,
between the models considered here, where the target space is spacetime itself, and those where the target space is another manifold,
as in nonlinear sigma models,
or a copy of the same manifold.
In all cases, when the target space metric
has Minkowski signature $-+++$,
the scalar associated to the time coordinate has a kinetic term
with opposite sign of those associated to the space coordinates.
Assuming that the sign of the Lagrangian is such that the
latter have the correct dynamics,
the time-like scalar will be a ghost.

As already mentioned,
in the context of massive gravity, two different strategies have been developed in order to avoid ghost instabilities:
the dRGT models, that preserve spacetime covariance,
and Lorentz-breaking models.
Both strategies can be used also for our models.
	
In the context of dRGT models of massive gravity,
it has been observed that certain nonlinear sigma models
with Minkowskian target space and special
actions built out of $\Tr(\sqrt{B})^n$
will be ghost-free \cite{deRham:2015ijs}.
By expanding $B = 1+X$ in terms of $X$, one can write 
$\Tr(\sqrt{B})^n$ as an infinite series of $\Tr B^n$. Hence, these actions can be viewed as special cases of our action
and the same construction can be applied also to our models.

In Lorentz-breaking massive gravity, one gives up Lorentz invariance and preserves only Euclidean symmetry of 3-dimensional space. Dubovsky and Rubakov \cite{Rubakov:2004eb, Dubovsky:2004sg} extended the non-linear St\"{u}ckelberg trick to Lorentz-violating massive gravity using the pullback of a 3-dimensional target space metric $h_{ab}$.
Similarly, we can construct ghost-free (but Lorentz-violating) dynamics of the diffeomorphisms of space, rather than spacetime. 
The target space metric $h(\varphi)$ would be only 3-dimensional 
in that case.
The solution $\varphi = Id_\Sigma$ is again a solution of the EOM
and it generates an effective cosmological term in the space directions.

\bigskip
\goodbreak

\section{Conclusions}

We have constructed a dynamical theory for diffeomorphisms
of the spacetime manifold,
which is closely related to models of dynamical coordinates.
From a geometrical point of view, coordinates are 
(locally defined) maps of spacetime
into a fixed Euclidean space, whereas diffeomorphisms
are maps of spacetime to itself.
Thus, dynamical diffeomorphisms naturally differ from dynamical
coordinates because of 
the identification of spacetime with the target space.

One motivation given in the introduction was that
in general, coordinates are only defined locally and
to cover a manifold with nontrivial topology,
several coordinate patches are needed. 
One could overcome this problem
by considering generalized models
with a target space that is homeomorphic to spacetime itself,
rather than flat Euclidean space.
One would then have a theory of diffeomorphisms from
one copy of spacetime into another.
Such models would still differ from the one considered here
as long as the target space metric is fixed.
The true difference lies in the identification 
of the spacetime and target space, and the respective metrics.

This lies at the root of all the other differences that
we have encountered.
First we have differences in the kinematics.
Suppose the dynamical coordinates are thought of as
scalar fields on spacetime.
Then diffeomorphisms act on them by right composition.
If the coordinates are viewed as a map into spacetime
(e.g. as fluid, analogous to comoving coordinates in cosmology),
diffeomorphisms act on them by left composition.
In both cases, the action of the group is free and transitive.
For a map of spacetime into itself, diffeomorphisms 
must act at both ends, resulting in an action by conjugation,
which is neither transitive nor free.

These kinematical differences inevitably give rise to
differences in the dynamics:
the energy-momentum tensor of diffeomorphisms
has an additional term that is not present in
the theories of dynamical coordinates,
and this term is conserved separately from the rest.
The identity is always a solution, but the
effective cosmological constant it generates is different
in the two cases.
For dynamical diffeomorphisms, it is given just by the term
that comes from the variation of $\sqrt{g}$ in the action.
\footnote{We observe here that in unimodular gravity
this contribution would be absent,
giving rise to a vanishing effective cosmological constant.
The resulting EM tensor is generally not conserved,
but as long as we are only interested in the identity
solution, it actually is.}

In this paper we have restricted our attention to the dynamics of the
diffeomorphisms by themselves.
We observe that if we allow the metric
to become dynamical, then the theory is no longer local:
the tensor $B$ of (\ref{defB}),
and consequently also the action (\ref{actgh}),
depends on the metric evaluated at two different points.
It would be interesting to investigate the possible couplings of
matter fields to diffeomorphisms.

Another motivation given in the introduction
was the issue of observables in gravity.
As in the case of Yang-Mills theories,
local, gauge-invariant expressions can be constructed
if a suitable ``compensator'' or ``St\"uckelberg'' field is present.
The configuration space of this field must be a copy
of the gauge group, and this is the case for our models,
but more is required:
the action of the gauge group on the compensator field must be free
and transitive.
To see this, it is enough to consider again the simple example of
a reference fluid, already mentioned in the introduction.
Let $F$ be a manifold labelling the fluid elements
and $\Phi:F\to M$ describe a configuration of the fluid.
\footnote{$\Phi$ is invertible and we can identify $\Phi^{-1}$
with the fields $\varphi$ of Appendix B.}
If $R:M\to \mathbb{R}$ is the Ricci scalar, for example,
$R\circ \Phi$ is the Ricci scalar evaluated in the reference
frame provided by the fluid.
This will be gauge invariant provided diffeomorphims of $M$
act on $\Phi$ by left composition: 
$R\circ \Phi\mapsto R\circ\psi\circ\psi^{-1}\circ\Phi$.
However, if we identify $F=M$, the diffeomorphisms of $M$
map act on $\Phi$ by conjugation, and 
$R\circ\Phi$ is no longer invariant,
being mapped to $R\circ\Psi\circ\psi^{-1}\circ\Phi\circ\psi$.
The lesson to be drawn is that in order to construct relational
gravitational observables, it is not useful to
consider diffeomorphisms of the spacetime manifold to itself.
One has to keep the domain and target space
separate, as in the models discussed in Appendix B.

We have limited ourselves here to general considerations.
Other aspects of the dynamics of diffeomorphisms,
especially the small fluctuations around the identity,
have been discussed in a framework that is very similar
to ours in \cite{Capoferri:2018fpp}.
For quantum aspects and the effective field theory description
of similar models we refer for example to
\cite{Leutwyler:1996er,
Endlich:2010hf,
Dubovsky:2011sj,
Ballesteros:2012kv}.

\section*{Acknowledgments}
We would like to thank M. Celoria, S. Gielen, P. A. H\"ohn, F. Nesti, M. Reuter for useful discussions and correspondence.

\vskip1cm
\goodbreak

\appendix

\section{Variation under infinitesimal diffeomorphisms}

In this appendix we prove the invariance of the action (\ref{actgh})
under infinitesimal diffeomorphisms.
There are variations of $g$ and variations of $\varphi$:
\bea
\delta_\xi S &=& \int dx \sqrt{g} \Bigg\{ \frac{1}{2}g^{\mu\nu} \delta_\xi g_{\mu\nu} \cL 
- \sum_n n\frac{\pa \cL}{\pa \tau_n}
(B^{n-1})^{\sigma \mu}
\delta_\xi g_{\mu \nu }
\nabla^\nu \vp^\gamma g_{\gamma\delta} (\vp)\nabla_\sigma\vp^\delta
\nn
&& \qquad \quad \qquad\qquad\qquad 
+\sum_n n\frac{\pa \cL}{\pa \tau_n} 
(B^{n-1})^\sigma{}_\mu g^{\mu \lambda}\nabla_\lambda \vp^\alpha 
\delta_\xi g_{\alpha \beta} (\vp)\nabla_\sigma \vp^\beta 
\nn
&& \qquad \quad \qquad\qquad\qquad 
+\sum_n n\frac{\pa \cL}{\pa \tau_n}\bigg(2 \nabla_\nu \delta_\xi \vp^\beta g_{\beta \alpha} (\vp)\nabla_\sigma \vp^\alpha (B^{n-1})^\sigma{}_\mu g^{\mu \nu} 
\nn
&&\qquad \quad \qquad\qquad\qquad 
+ (B^{n-1})^\sigma{}_\mu g^{\mu \nu} \nabla_\nu \vp^\beta \d_\xi\vp^\a \pa_\a g_{\beta \gamma}(\vp) \nabla_\sigma \vp^\gamma\bigg)
 \Bigg\}
\nn
&=&
\int dx \sqrt{g} \Bigg\{ -2 \xi_\nu\nabla_\mu
\Bigg[\frac{1}{2}g^{\mu\nu}\cL 
-\sum_n n\frac{\pa \cL}{\pa \tau_n}(B^{n-1})^{\sigma\mu} 
\nabla^\nu \vp^\gamma g_{\gamma\delta} (\vp) \nabla_\sigma\vp^\delta\Bigg] 
\nn
&&\qquad \quad \qquad\qquad\qquad 
 + 2 D_\a\xi_\b(\vp) \sum_n n\frac{\pa \cL}{\pa \tau_n}\nabla_\sigma \vp^\beta (B^{n-1})^\sigma{}_\mu g^{\mu \lambda}\nabla_\lambda \vp^\alpha \nn
 &&+ \xi^\tau\nabla_\tau\vp^\a \Bigg[-\nabla_\mu\bigg(2\sum_n n\frac{\pa \cL}{\pa \tau_n} (B^{n-1})^\mu{}_\rho g^{\rho \tau}\nabla_\tau \vp^\beta g_{\beta \alpha}(\vp) \bigg) \nn
 &&
 \qquad \quad \qquad\qquad\qquad + \sum_n n\frac{\pa \cL}{\pa \tau_n} (B^{n-1})^\mu{}_\rho g^{\rho\tau} \pa_\tau\vp^\beta 
\pa_\a g_{\b\gamma}(\vp) \pa_\mu \vp^\gamma\Bigg]\nn
 &&
 - \xi^\a(\vp)\Bigg[-\nabla_\mu\bigg(2\sum_n n\frac{\pa \cL}{\pa \tau_n} (B^{n-1})^\mu{}_\rho g^{\rho \tau}\nabla_\tau \vp^\beta g_{\beta \alpha}(\vp) \bigg) \nn
&&
\qquad \quad \qquad\qquad\qquad 
+ \sum_n n\frac{\pa \cL}{\pa \tau_n} 
(B^{n-1})^\mu{}_\rho g^{\rho\tau} 
\pa_\tau\vp^\beta \pa_\a g_{\b\gamma}(\vp)\pa_\mu\vp^\gamma\Bigg]
 \Bigg\}\ .
 \label{paola}
\eea
The first, third and fourth lines are the same as one would have in an ordinary nonlinear sigma model in curved space, where the target space metric is not affected by spacetime diffeomorphisms.
By explicitly taking the derivatives of the first line,
one finds that they cancel the third and the fourth line.

The second, fifth and sixth lines are novel.
They are characterized by the fact that the
infinitesimal parameter $\xi$ is evaluated at $\varphi(x)$.
The second line can be expanded as follows:
$$ 
2(\partial_\alpha\xi_\beta(\varphi)-\Gamma_{\alpha\beta}^\lambda(\varphi)
\xi_\lambda(\varphi))
\sum_n n\frac{\pa \cL}{\pa \tau_n} 
(B^{n-1})^{\sigma\lambda}\nabla_\lambda \vp^\alpha
\nabla_\sigma \vp^\beta\ .
$$
As recalled earlier, $\xi_\lambda(\varphi)$ is a scalar under diffeomorphisms,
so by the chain rule its covariant derivative is
$\partial_\lambda\varphi^\alpha
\partial_\alpha\xi_\beta(\varphi)
=\nabla_\lambda\xi_\beta(\varphi)$.
Therefore the first term can be integrated by parts and we obtain
\be
-2\xi_\beta(\varphi)
\nabla_\lambda\left(
\sum_n n\frac{\pa \cL}{\pa \tau_n} 
(B^{n-1})^{\sigma\lambda}
\nabla_\sigma \vp^\beta\right)
-2\Gamma_{\alpha\beta}^\lambda(\varphi)\xi_\lambda(\varphi)
\sum_n n\frac{\pa \cL}{\pa \tau_n} 
(B^{n-1})^{\sigma\lambda}\nabla_\lambda \vp^\alpha
\nabla_\sigma \vp^\beta
\label{elena}
\ee
Next, in the fifth line we separate the covariant derivative
acting on $g_{\alpha\beta}(\varphi)$ from the rest,
and we get
\be
2\xi_\beta(\varphi)
\nabla_\mu\left(
\sum_n n\frac{\pa \cL}{\pa \tau_n} 
(B^{n-1})^{\mu\tau}
\nabla_\tau \vp^\beta\right)
+2\xi^\alpha(\varphi)
\sum_n n\frac{\pa \cL}{\pa \tau_n} 
(B^{n-1})^{\mu\tau}\nabla_\mu \vp^\gamma
\nabla_\tau \vp^\beta
\partial_\gamma g_{\alpha\beta}(\varphi)
\ee
The first term cancels the first term of (\ref{elena}).
The second term combines with the sixth line
to reconstruct a Christoffel symbol,
and the result cancels with the second term in (\ref{elena}).
In this way also these terms cancel out,
and we have proven the invariance of the action.

\medskip

We can now also see the differential identities that follows from
$\diffM$-invariance.
To this end, we have to work a little more on the 
second line of (\ref{paola}).
Namely, as in the derivation of the EM tensor,
we want to have the infinitesimal parameter $\xi$
evaluated at $x$ rather than $\varphi(x)$.
This is dealt with by the method already used in Sect. 2.2,
namely changing coordinates from $x$ to $x'=\varphi(x)$.
Then the second line becomes
$$
\int dx\sqrt{g'} \sum_n n\left( \frac{\pa \cL}{\pa \tau_n} \right)' (B'^{n-1})^{\alpha\beta}\nabla_\alpha\xi_\beta
=-\int dx\sqrt{g}\,\xi_\beta\nabla_\alpha\left(
\frac{\sqrt{g'}}{\sqrt{g}} \sum_n n\left( \frac{\pa \cL}{\pa \tau_n} \right)' (B'^{n-1})^{\alpha\beta}\right)\ .
$$
Note that $\nabla$ is the connection obtained from the 
metric $g$, so $\nabla_\rho g_{\mu\nu}=0$,
but $\nabla_\rho g'_{\mu\nu}$ is not zero in general.
We recognize that the content of the round bracket
is the tensor $T_{(L)}^{\mu\nu}$ consisting of
the last term in (\ref{emtensor}),
whereas the round bracket in the first line of (\ref{paola}),
is just the tensor $T_{(R)}^{\mu\nu}$
consisiting of the first two terms of (\ref{emtensor}).
On the other hand, the expression in square brackets
in the third and fourth lines, and in the fifth and sixth lines
is nothing but the EOM.
Therefore (\ref{paola}) is just (\ref{diffinv}) written more explicitly, and where we recognize the separate
conservation of $T_{(R)}^{\mu\nu}$ and $T_{(L)}^{\mu\nu}$.

\section{Models with different domain and target}

In this appendix we discuss models of dynamical diffeomorphisms where,
in contrast to the models discussed in the main text,
the domain $M_R$ and the target space $M_L$ are viewed as different manifolds, and consequently also their metrics are different.
Of course since the manifolds are diffeomorphic one can also
view them as ``the same manifold'',
but this presupposes a preferred identification,
whereas here we will not assume one, at least not to begin with.

The formalism lends itself to two rather different interpretations.
In the ``field theoretic'' interpretation,
$M_R$ is spacetime and $M_L$ is some internal space,
endowed with a fixed metric.
In the ``brane'' interpretation, $M_R$ is the brane worldsheet
and $M_L$ is spacetime.

\subsection{Left and right diffeomorphisms}

We denote $\diff$ the space of diffeomorphisms of $M_R$ to $M_L$.
The diffeomorphisms of $M_R$ and $M_L$ into themselves will
be denoted $\diffr$ and $\diffl$ respectively.
They act in the usual way on tensors on $M_R$ and $M_L$
respectively.
The groups $\diffr$ and $\diffl$ act on $\diff$
by right and left composition
\bea
\varphi\mapsto\varphi'&=&\varphi\circ\psi_R\ ,
\nonumber
\\
\varphi\mapsto\varphi'&=&\psi_L^{-1}\circ\varphi\ .
\eea
Each of these actions is free and transitive.
In particular, $\diffl\times\diffr$ 
acts transitively on $\diff$,
so we can transform any $\varphi\in\diff$ to 
any other $\varphi'$.
Now pick some fixed $\bar\varphi\in\diff$.
It defines an isomorphism $\iota:\diffr\to\diffl$ by
\be
\iota(\psi_R)=\bar\varphi\circ\psi_R\circ\bar\varphi^{-1}\ .
\label{iota}
\ee
The stabilizer of $\bar\varphi$ is the ``diagonal'' subgroup
$\Delta\diff$ consisting of transformations
$(\psi_L,\psi_R)\in\diffr\times\diffl$
of the form
$$
(\psi_L,\psi_R)=(\iota(\psi),\psi)\ .
$$
The diagonal subgroup acts on $\diff$ as follows:
for $f\in\diffr$,
\be
f:\varphi\mapsto
\iota(f^{-1})\circ\varphi\circ f
=\bar\varphi\circ f^{-1}\circ\bar\varphi^{-1}
\circ\varphi\circ f
\label{pia}
\ee
and indeed under such action $\bar\varphi$ is invariant
$$
f:\bar\varphi\mapsto 
\bar\varphi\circ f^{-1}\circ\bar\varphi^{-1}\circ\bar\varphi\circ f
=\bar\varphi\ .
$$
We have shown that the configuration space $\diff$ can be regarded
as a homogeneous space
\be
(\diffr\times\diffl)/\Delta\diff\ .
\label{dddd}
\ee
All this bears a striking similarity to chiral models of particle physics, but here the groups are infinite dimensional.

\subsection{Dynamics}

Let us denote $g$ and $h$ the metrics in $M_R$ and $M_L$,
respectively.
The actions we are interested in have the form
\be
S=\int_{M_R} dx\,\sqrt{g}\,\cL(\sigma_1,\sigma_2,\sigma_3,\sigma_4)\ ,
\label{actB}
\ee
where $\sigma_n=\tr B^n$ and $B^\mu{}_\nu$
is given by (\ref{B1}).
The usual action for nonlinear sigma models corresponds to
$\cL=-\frac12\sigma_1$ and one may keep this example in mind
in the following.

As long as the metrics $g$ and $h$ are kept fixed,
the action has, generically, no symmetries.
In the field theoretic interpretation 
the target space metric $h$ can be interpreted as an infinite
set of coupling constants. Symmetries of a theory correspond to
transformations that act on the dynamical variables
leaving the couplings fixed, so left diffeomorphisms 
are not symmetries, unless $h$ has some isometries.
The proper interpretation of left (target space) diffeomorphisms
is as field redefinitions.
In the brane interpretation, $h$ represents the spacetime metric
and its dynamics, in the quantum theory,
comes from the beta functions of the worldsheet quantum field theory.
Once again, left (spacetime) diffeomorphisms cannot be
interpreted as symmetries.
We shall use the word ``invariances''
for transformations that leave the action invariant,
from a mere mathematical viewpoint,
and irrespective of their physical interpretation.

With these cautionary remarks in mind,
the action is separately invariant under $\diffr$ and $\diffl$,
as we shall now see.
Under a right-diffeomorphism the metric $g$ is pulled back
but the metric $h$ is invariant. 
The pullback $\varphi^*h$ transforms like an ordinary tensor on $M_R$:
$$
\varphi^{\prime*} h'
=(\varphi\circ\psi_R)^* h=
\psi_R^*(\varphi^*h)\ .
$$
Since the integrand of the action is a scalar density on $M_R$,
the action is $\diffr$-invariant.

On the other hand under a left-diffeomorphism $h$ gets pulled back
but $g$ is invariant.
The pullback $\varphi^*h$ is invariant, because:
$$
\varphi^{\prime*} h'
=(\psi_L^{-1}\circ\varphi)^* (\psi_L^*h)
=\varphi^*\circ \psi_L^{-1*}\circ\psi_L^*h
=\varphi^*h\ .
$$
Since $g$ is also invariant, the action is 
trivially $\diffl$-invariant.

The equation of motion of these models has the form (\ref{eom}),
except that $B$ is given by (\ref{B1}) instead of (\ref{defB}),
i.e. $g_{\alpha\beta}(\varphi)$ is replaced
everywhere by $h_{\alpha\beta}(\varphi)$
and the Christoffel symbol appearing in (\ref{defD})
is the Christoffel symbol of $h_{\alpha\beta}$
(whereas $\nabla_\mu$ in (\ref{defnabla})
is still constructed with the
Christoffel symbols of $g_{\mu\nu}$).
The EM tensor defined by (\ref{enmom})
is equal to (\ref{emtensor}), but the last term is absent. 
It thus agrees with what we called
$T_{(R)}^{\mu\nu}$.
Separately, we define the tensor
\be
T_{(L)}^{\alpha\beta}(y)=\frac{2}{\sqrt{h(y)}}
\frac{\delta S}{\delta h_{\alpha\beta}(y)}\ .
\label{enmomL}
\ee

In the next subsections we consider the infinitesimal versions
of these transformations
and the differential identities that follow
from the invariances of the action.

\subsection{Consequences of right diffeomorphism invariance}

A right diffeomorphism acts on points of $M_R$ by
$x\mapsto x'=\psi^{-1}(x)$.
The infinitesimal version is
$$
\delta x^\alpha=-\xi^\alpha(x)\ .
$$
The variation of any tensor $T$ on $M_R$ is its Lie derivative
$\delta_\xi T={\cal L}_\xi T$.
For the metric
$$
\delta_\xi g_{\mu\nu}
=\xi^\rho\partial_\rho g_{\mu\nu}
+g_{\mu\rho}\partial_\nu\xi^\rho
+g_{\rho\nu}\partial_\mu\xi^\rho
=\nabla_\mu\xi_\nu+\nabla_\nu\xi_\mu\ .
$$
The infinitesimal variation of $\varphi$ is
\be
\delta_\xi\varphi^\rho(x)
=\xi^\lambda\partial_\lambda\varphi^\rho
\ee
and $\delta_\xi h_{\alpha\beta}=0$.
Now varying the pullback we have
\bea
\delta_\xi (\varphi^*h_{\mu\nu})
&=&\delta_\xi 
(\partial_\mu\varphi^\alpha\partial_\nu\varphi^\beta h_{\alpha\beta}(\varphi))
\nonumber\\
&=&
\partial_\mu\delta\varphi^\alpha\partial_\nu\varphi^\beta h_{\alpha\beta}(\varphi)+
\partial_\mu\varphi^\alpha\partial_\nu\delta\varphi^\beta h_{\alpha\beta}(\varphi)
+\partial_\mu\varphi^\alpha\partial_\nu\varphi^\beta 
\partial_\gamma h_{\alpha\beta}(\varphi)\delta\varphi^\gamma\ .
\nonumber
\eea
Inserting the above formulae for the variation
and expanding, one arrives after a few steps at
\bea
\xi^\rho\partial_\rho (\varphi^*h)_{\mu\nu}
+(\varphi^*h)_{\mu\rho}\partial_\nu\xi^\rho
+(\varphi^*h)_{\rho\nu}\partial_\mu\xi^\rho
={\cal L}_\xi (\varphi^*h)_{\mu\nu}\ .
\eea
which just confirms that $\varphi^*h$ transforms as a tensor.

As usual, from the diffeomorphism invariance
one can obtain a differential identity.
The derivation follows the steps of Appendix A,
but with some terms now absent.
In the end one obtains
\be
0=\int d^4x\sqrt{g}
\left[\xi^\tau\pa_\tau\vp^\alpha E_\alpha
-\xi_\mu\nabla_\nu T_{(R)}^{\mu\nu}
\right]\ .
\ee
In contrast to (\ref{diffinv}), 
since $\pa_\tau\varphi^\alpha$ is nondegenerate,
the coefficient of the EOM is always nonzero
and therefore we find that the EOM and EM conservation
are completely equivalent.
This is a consequence of the action of the group
being free and transitive.

\subsection{Consequences of left diffeomorphism invariance}

A left diffeomorphism acts on points of $M_L$ by
$x\mapsto x'=x$ and $\varphi(x)\mapsto \psi^{-1}(\varphi(x))$.
The infinitesimal versions are
$$
\delta_\xi x^\mu=0\ ,\quad
\delta_\xi \varphi^\alpha(x)=-\xi^\alpha(\varphi(x))\ .
$$

The variation of any tensor $T$ on $M_L$ is its Lie derivative
$\delta_\xi T={\cal L}_\xi T$,
and tensors on $M_R$ are invariant.
For the metric in $M_L$
\be
\delta_\xi h_{\alpha\beta}
=\xi^\rho\partial_\rho h_{\alpha\beta}
+h_{\alpha\rho}\partial_\beta\xi^\rho
+h_{\rho\beta}\partial_\alpha\xi^\rho
=D_\alpha\xi_\beta+D_\beta\xi_\alpha\ ,
\label{alberto}
\ee
where we used the notation (\ref{ettore}).
Now varying the pullback and using this formula we have
\bea
\delta_\xi (\varphi^*h_{\mu\nu})
&=&\delta_\xi 
(\partial_\mu\varphi^\alpha\partial_\nu\varphi^\beta h_{\alpha\beta}(\varphi))
\nonumber\\
&=&
\partial_\mu\delta_\xi\varphi^\alpha\partial_\nu\varphi^\beta h_{\alpha\beta}(\varphi)+
\partial_\mu\varphi^\alpha\partial_\nu\delta_\xi\varphi^\beta h_{\alpha\beta}(\varphi)
\nonumber\\
&&
+\partial_\mu\varphi^\alpha\partial_\nu\varphi^\beta 
\delta_\xi\varphi^\gamma\partial_\gamma h_{\alpha\beta}(\varphi)
+\partial_\mu\varphi^\alpha\partial_\nu\varphi^\beta 
\delta_\xi h_{\alpha\beta}(\varphi)
\nonumber\\
&=&0\ .
\eea
As we have already seen at the level of finite transformations,
both $g$ and $\varphi^*h$ are invariant,
and therefore the invariance of the action is trivial.

Since the metric $g$ is unaffected by these
transformations, no consequence can be derived from
$\diffl$-invariance concerning the energy-momentum tensor
$T_{(R)}^{\mu\nu}$.
Nevertheless, we can obtain another differential identity
involving the tensor (\ref{enmomL}).

Since in the action the metric $h$ always appears evaluated at 
$\varphi(x)$, it is convenient to change integration variable
from $x$ to $x'=\varphi(x)$, and write
$$
S=\int dx'\sqrt{g'(x')}\cL'(x')\ .
$$
Since $\cL$ is a given function of the trace invariants
$\cL(x)=F(\tau_1(x),\tau_2(x),\tau_3(x),\tau_4(x))$,
the transformed Lagrangian $\cL'$ will be the same function of the transformed invariants
$\cL'(x')=F(\tau'_1(x'),\tau'_2(x'),\tau'_3(x'),\tau'_4(x'))$
and since $\tau'_n(x')=\tau_n(x)$, also  $\cL'(x')=\cL(x)$.

For example, for the Lagrangian (\ref{lag1}),
we can write
$$
\cL'(x')=-\frac12 (\varphi_* g^{-1})^{\alpha\beta}(x')
h_{\alpha\beta}(x')
$$
and therefore
$$
T_{(L)}^{\alpha\beta}(y)=
-f^2\frac{\sqrt{g'}}{\sqrt{h}}g^{\prime\alpha\beta}(y)\ ,
$$
where we write $g^{\prime\alpha\beta}$ for
the push-forward of the inverse metric, 
$(\varphi_* g^{-1})^{\alpha\beta}(x')$.
In general
\be
T_{(L)}^{\alpha\beta}(y)=
2\frac{\sqrt{g'}}{\sqrt{h}}
\sum_{n=1}^4 n 
\left(\frac{\partial\cL}{\partial\tau_n}\right)^\prime
(B^{\prime n-1})^{\alpha\beta}(y)\ ,
\ee

Using the infinitesimal variation (\ref{alberto})
and the invariance of the action under left diffeomorphisms,
one finds that

\be
0=\delta_\xi S
=\int dx'\sqrt{h(x')}\left[
-\xi^\alpha(x') E_\alpha(x')
+\xi_\beta(x')D_\alpha T_{(L)}^{\alpha\beta}(x')\right]
\ee
and therefore, on shell
\be
D_\alpha T_{(L)}^{\alpha\beta}=0\ .
\ee

\subsection{Relation to the models in the main text}

Even though the manifolds $M_R$ and $M_L$ are,
by assumption, diffeomorphic,
there will in general be no relation between the
respective metrics.
Consider the special case in which there exists
a diffeomorphism $\bar\varphi$ such that:
\be
\bar\varphi^* h=g\ .
\ee
If $h$ and $g$ have isometries, $\bar\varphi$ will not
be unique. We disregard this case here.
We can use $\bar\varphi$ to define a preferred identification 
of $M_R$ and $M_L$ and, via equation (\ref{iota}), a preferred identification of the respective diffeomorphism groups.

Since $\bar\varphi$ is a diffeomorphism,
without loss of generality we can choose atlases on $M_L$
and $M_R$ to be related by $\bar\varphi$.
This means that on any chart, if $x^\alpha$ are the coordinates of a 
point $x$ and $y^\alpha$ are the coordinates of $\bar\varphi(x)$,
then
\be
y^\alpha=x^\alpha\ .
\ee

If we use $\bar\varphi$ to identify $M_R$ and $M_L$,
we have only one manifold $M$,
$\bar\varphi$ can be thought of as the
identity mapping of $M$ to itself
and the action (\ref{pia}) becomes conjugation.
Since $\bar\varphi$ is now a fixed element of the theory,
the original invariance under 
$\diffr\times\diffl$ is broken to $\Delta\diff$,
acting by conjugation.
In this way we recover the models of the main text.

\vskip 1cm


\end{document}